\input harvmac
\input epsf


\def\h{{1\over2}}
\def\p{\partial}
\def\vp{\varphi}
\def\ap{\alpha}

\Title{}{\vbox{\centerline{Nonminimal Inflation and the Running Spectral
Index}}} \centerline{Miao Li$^{1,2}$}
\medskip
\centerline{\it $^1$ Interdisciplinary Center of Theoretical
Studies} \centerline{\it USTC, Hefei, Anhui, China}
\centerline{\it and}
\medskip
\centerline{\it $^2$ Institute of Theoretical Physics}
\centerline{\it Academia Sinica, P. O. Box 2735} \centerline{\it
Beijing 100080, China}
\medskip

\centerline{\tt mli@itp.ac.cn}
\medskip

We study a class of models in which the inflaton is minimally coupled to gravity with a
term $f(R)\vp^2$. We focus in particular on the case when $f\sim R^2$, the expansion
of the scale factor is driven by the usual potential energy, while the rolling of the inflaton
is driven by the nonminial coupling. We show that the power spectrum is in general blue,
and the problem of getting a running spectral index is eased. However, the inflaton potential
must have a large second derivative in order to get a large running.

\Date{July 2006}

\nref\wamp{D. N. Spergel et al., astro-ph/0603449.}
\nref\hl{Q.-G. Huang and M, Li, astro-ph/0603782.}
\nref\zhang{X. Zhang and F.-Q, Wu, astro-ph/0604195, Phys. Lett.
B638 (2006) 396; Q.-G. Huang, astro-ph/0605442.}
\nref\hlf{Q.G. Huang and M. Li, JHEP 0306(2003)014, hep-th/0304203; JCAP 0311(2003)001,
astro-ph/0308458; Nucl.Phys.B 713(2005)219,
astro-ph/0311378; S. Tsujikawa, R. Maartens and R. Brandenberger,
astro-ph/0308169.}
\nref\prunt{K. Izawa, hep-ph/0305286; S. Cremonini, hep-th/0305244;
E. Keski-Vakkuri, M. Sloth, hep-th/0306070; M. Yamaguchi, J.
Yokoyama, hep-ph/0307373; J. Martin, C. Ringeval, astro-ph/0310382;
K. Ke, hep-th/0312013; S. Tsujikawa, A. Liddle, astro-ph/0312162; C.
Chen, B. Feng, X. Wang, Z. Yang, astro-ph/0404419; L. Sriramkumar,
T. Padmanabhan, gr-qc/0408034; N. Kogo, M. Sasaki, J. Yokoyama,
astro-ph/0409052; G. Calcagni, A. Liddle, E. Ramirez,
astro-ph/0506558; H. Kim, J. Yee, C. Rim, gr-qc/0506122. }

\nref\pruns{A. Ashoorioon, J. Hovdebo, R. Mann, gr-qc/0504135; G.
Ballesteros, J. Casas, J. Espinosa, hep-ph/0601134; J. Cline, L.
Hoi, astro-ph/0603403. }

\nref\run{B. Feng, M. Li, R. J. Zhang and X. Zhang,
astro-ph/0302479; J. E. Lidsey and R. Tavakol, astro-ph/0304113; L.
Pogosian, S. -H. Henry Tye, I. Wasserman and M. Wyman, Phys. Rev. D
68 (2003) 023506, hep-th/0304188; S. A. Pavluchenko,
astro-ph/0309834; G. Dvali and S. Kachru, hep-ph/0310244. }
\nref\alnv{J. C. Hwang, ``Cosmological perturbations in generalized gravity
theories: Formulation," Class. Quant. Grav. 7 (1990) 1613;
J. C.  Hwang and H. Noh, ``Classical evolution and quantum generation in generalized gravity
theories including string corrections and tachyon: Unified analyses," gr-qc/0412126,
Phys. Rev. D71(2005)063536.}
\nref\lww{B. Chen, M. Li, T. Wang and Y. Wang, work in progress.}

The WMAP three data reveal that the power spectrum of the primordial density fluctuation
is not exactly scale invariant \wamp. If we do not assume the running of the spectral
index, $n_s\simeq 0.95$. However, once the running of the spectral index is introduced
as a parameter, the data favor a blue spectrum with a large running. It was shown that
the noncommutative inflation model can in principle account for this large running \hl,
but with a rather unnatural potential \zhang. For earlier studies on inflation models
with a large running, see \refs{\hlf-\run}.

We shall study an alternative class of models with the inflaton nonminimally coupled to
gravity. This class of models is among those studied in \alnv, however, those authors
did not study concrete models and moreover their method of treating fluctuation
equations is limited to cases not including the class of models studied in this note.

In the following, we shall assume a number of simplifications, leaving a more careful
and detailed study to \lww. We shall see that when the nonminimal coupling term
$f(R)\vp^2$ assumes the form $f(R)\sim R^2$ and inflation is slow-rolling, the evolution
of the scale factor is driven by the potential energy $V(\vp)$ only. There is a region
in the parameter space where the rolling of the inflaton is driven by only the nonminimal
coupling term, we shall focus on this region, and once again leaving the study of more general
situations to \lww.

\bigskip

Consider the following Einstein-Hilbert action coupled to a nonminimal inflaton
\eqn\ehat{S=\h M_p^2\int d^4x\sqrt{-g}R+\int d^4x\sqrt{-g}\left(-\h (\nabla \vp)^2-V(\vp)
-\h f({1\over 6}R)\vp^2\right),}
where $M_p^2={1\over 8\pi G}$ is the reduced Planck mass squared, $f$ is a
function of the scalar curvature $R$. We will work with the FRW metric
\eqn\frwm{ds^2=-dt^2+a^2(t)dx^2=a^2(\tau)(-d\tau^2+dx^2),}
where $t$ is the co-moving time and $\tau$ is the conformal time. The scalar curvature
of the FRW metric reads:
\eqn\frwc{R=6\left({\ddot{a}\over a}+({\dot{a}\over a})^2\right)=6{a''\over a^3}.}
where the dot denotes the derivative with respect to $t$ and the prime denotes the derivative
with respect to $\tau$.
The equations of motion derived from \ehat\ are
\eqn\fes{\eqalign{3M_p^2a'^2&=\h a^2\vp'^2+(V+\h f \vp^2)a^4-f'aa''\vp^2+\h\p_\tau(f'aa'\vp^2),\cr
6M_p^2a''&=-a\vp'^2+(4V+2f\vp^2)a^3-{3\over 2}f'a''\vp^2+\h \p^2_\tau(f'a\vp^2),\cr
\vp''+&2(a'/a) \vp' +(V'+f\vp)a^2=0,}}
note that $f'$ is the derivative of $f$ with respect to its argument ${1\over 6}R$, and
$V'$ is the derivative of $V$ with respect to $\vp$
.

We shall impose the usual slow-roll conditions:
\eqn\src{|\ddot{\vp}\ll 3H|\dot{\vp}|,\quad |\dot{H}|\ll H^2.}
In order to simplify calculations, we assume $\h (\dot{\vp})^2\ll V$, the above slow-roll
conditions lead to
\eqn\srcn{\eqalign{|V''+f-{12f'H^2\dot{H}\vp\over V'+f\vp}|&\ll 9H^2,\cr
|f'\dot{H}\vp^2|&\ll V.}}
To make the maximal use of  the nonminimal coupling term $f\vp^2$, we shall assume that in the
e.o.m. for $\vp$, the force due to the nonminimal coupling is much larger than the one
caused by the usual potential $V$, namely
\eqn\bass{|f\vp|\gg |V'|.}
With this assumption, the third term on the L.H.S. of the first equation in \srcn\ becomes
$${12f'H^2\dot{H}\over f}$$
and it is much smaller than $12H^2$ parametrically if $f$ is a monomial of $R$. In this case,
\srcn\ simplifies to
\eqn\srcnn{|V''+f|\ll 9H^2,\quad |f'\dot{H}\phi^2| \ll V.}
With these conditions in mind, after some lengthy calculations, we find the much
simplified equations of motion
\eqn\sseom{\eqalign{3M_p^2H^2&=\h (f-f'H^2)\vp^2 +V,\cr
3H\dot{\vp}&=-f\vp.}}

Let us now specify to a class of models with $f={1\over 4}l^2({1\over 6}R)^2$, where $l$ is a
length scale being either the string scale or the Planck scale. With $R=12H^2$,
$f=l^2H^4$, with this choice, the first term $(f-f'H^2)\vp^2$ in the Friedman equation vanishes
identically. Thus, the inflation of the scale factor is driven by the potential energy $V$
only, and with assumption \bass, the rolling of the inflaton is driven by $f\vp$ only.
Define slow-roll parameters
\eqn\slpa{\epsilon={M_p^2\over 2}({V'\over V})^2,\quad \eta =M_p^2{V''\over V}, \quad
\Delta ={l^2V\over 9M_p^2},}
where $\epsilon$ and $\eta$ are the traditional ones, $\Delta$ is a new slow-roll parameter
in our nonminimal models. Now, if $|\eta|\ll 1$ and $|\Delta|\ll 3$, the first slow-roll
condition in \srcnn\ is met. The dominance condition \bass\ becomes
\eqn\dcc{\Delta {\vp^2\over M_p^2}\gg |{V'\over V}\vp|.}
If ${V'\over V}\vp\sim 1$, the above condition implies $\vp^2\gg M_p^2$, we must have
large field inflation. To deal with the second slow-roll condition in \srcnn, we use the Friedman
equation to compute $\dot{H}$ and in the end we obtain
\eqn\nsc{\Delta^2{\vp^2\over M_p^2}\ll 2|{V\over V'\vp}|.}
When we consider concrete models, we need to come back to conditions \dcc\ and \nsc.

Let $\vp_*$ be the value of $\vp$ when inflation ends, the number of e-folds is
\eqn\numef{N=\int Hdt=-{1\over M_p^2}\int^{\vp_*}_\vp d\vp {V\over f\vp}=-
\int^{\vp_*}_\vp {d\vp\over \Delta \vp}.}
It is clear that to have a large e-folds number, $\Delta$ must be small. For positive $\vp$,
$\vp$ always rolls from larger values to smaller ones, and when the slow-roll conditions
are violated, inflation ends.

To calculate fluctuations, we need to perturb the Einstein
equations. As usual, we will work with the following perturbed
metric
\eqn\ptme{ds^2=a^2\left(-(1+2\phi)d\tau^2+(1-2\psi)dx^idx^i\right).}
The perturbation of $\vp$ is denoted by $\delta\vp$. What $\phi,
\psi$ and $\delta\vp$ represent is called scalar perturbation, these
components are not independent. In the usual minimal inflation
models, there are the following linear relations
\eqn\lrelt{\psi'+{a'\over a}\phi=\h M_p^{-2}\vp'\delta\vp,\quad
\phi=\psi.} These relations are modified for the nonminimal models.
However, for simplicity, we will still use the above relations, a
honest treatment can be found in \lww. Let $u=a\delta\vp$, the
perturbation equations are
\eqn\peom{\eqalign{&\phi''-\Delta\phi+(2\ap-{2\vp''\over\vp'})\phi'+(2\ap'-2\ap{\vp''
\over\vp'})\phi=0,\cr &u''-\Delta
u+(-\ap'-\ap^2+V''+f-2M_p^{-2}\vp'^2)u-2a\vp''\phi=0,}} where
$\ap=a'/a$. We are interested in the power spectrum of the curvature
perturbation ${\cal R}=\phi+{a\over z}\delta\vp$, where
$z=a\dot{\vp}/H$. However, the e.o.m for ${\cal R}$ is not
diagonalized. Instead, we will use $\Phi=z{\cal R}$, the e.o.m. for
$\Phi$ read \eqn\eomp{\Phi''-\Delta \Phi-{z''\over z}\Phi+hu=0,}
where \eqn\hgd{\eqalign{h&={2g\over
\ap}({\vp''\over\vp'}+2\ap-{\ap'\over\ap})+{g'\over\ap} +\p_\tau f
{a^2\vp\over\vp'},\cr g&={1\over 6}M_p^{-2}f'(\ap^2-\ap')\vp^2.}}
Thus, the e.o.m. for $\Phi$ is not in a diagonal form. Nevertheless,
we can prove that \eqn\cons{g\ll {1\over 6}\epsilon a^2H^2,\quad
h\leq \epsilon a^2H^2,} so to the first order in $\epsilon$, one can
drop the last term in \eomp. Moreover, as far as the slow-roll
conditions are met, the solution to equation \eomp, to the first
order, is the same as in a standard inflation model. The power
spectrum, to the first order, is given by
\eqn\psp{\delta_H^2={1\over 25\pi^2}{H^2\over\dot{\vp}^2}={3\over
25\pi^2l^2}{1\over \Delta \vp^2}.} A more careful calculation of the
power spectrum can be found in \lww. Now, although $\Delta$ is small,
it is possible to choose appropriate $l$ and $\vp$ to make $\delta_H^2$ sufficiently
small to match the COBE normalization ($\sim 4\times 10^{-10}$).
For instance, take $\Delta=10^{-1}$, this demands $l\vp\sim 10^4$, this is
easily satisfied if $l$ is three magnitudes larger than $l_p$ and $\vp$
is one magnitude larger than $M_p$.

To compute the power spectral index $n_s$ and its running $\ap_s=dn_s/d\ln k$,
we use the old horizon exit condition $k=aH$, thus
\eqn\hexi{d\ln k=\h d\ln V+{d\ln a\over d\vp}d\vp=
-(\Delta^{-1}-\h {V'\vp\over V}){d\vp\over\vp}.}
Since $\Delta\sim V$, we can also write
\eqn\hexit{d\ln k=-(\Delta^{-1}-\h{\Delta'\vp\over\Delta}){d\vp\over\vp},}
where $\Delta'=d\Delta/d\vp$. Therefore,
\eqn\spind{n_s-1=(2+{\Delta'\vp\over\Delta})(1-\h\Delta'\vp)^{-1}\Delta,}
and
\eqn\indr{\eqalign{\ap_s&=[{\vp^2\over M_p^2}(\eta-2\epsilon)+{\Delta'\vp\over\Delta}]
(1-\h\Delta'\vp)^{-2}\Delta^2\cr
&+(2+{\Delta'\vp\over\Delta})[{\vp^2\over M_p^2}(\eta-2\epsilon)\Delta +(-1+\Delta)
{\Delta'\vp\over\Delta}](1-\h\Delta'\vp)^{-3}\Delta^2.}}

To examine the consequences of formulas \spind\ and \indr, let us consider a further
simplification: Since we assumed $\Delta\ll 1$, as required by the slow-roll conditions,
it is also natural to assume $\Delta'\vp\ll 1$. We have a simpler formula for the
spectral index
\eqn\sspind{n_s-1=(2+{\Delta'\vp\over\Delta})\Delta.}
Neglecting some terms in the second line of \indr\ smaller by a factor $\Delta$ compared with
the first line of \indr, we have
\eqn\srin{\ap_s=-({\vp^2\over M_p^2}\eta +3{\Delta'\vp\over\Delta})\Delta^2.}
The above formula can also be derived directly from \sspind\ using $d\ln k
=-\Delta^{-1}d\vp/\vp$, since $(\vp^2/M_p^2)\eta$ is just $\vp^2\Delta''/\Delta$.

$\bullet$ Large running

The WMAP3 result indicates that if we introduce the running of the spectral index,
the running is quite large. At $k=0.05$Mpc$^{-1}$, $n_s=1.21$, $\ap_s=-0.1$, we thus
have $(n_s-1)^2/|\ap_s|=0.44$, a value smaller than 1. In the usual slow-roll inflation
models, this ratio is generally greater than 1, thus we need to resort to new models to
account for this small ratio. Another difficulty in the usual slow-roll models is that
$n_s-1$ is often negative, namely the power spectrum is red.

To match the running data, we require
\eqn\consd{2+{\Delta'\vp\over \Delta}>0,\quad {\Delta''\vp^2\over\Delta} +3{\Delta'\vp
\over \Delta} >0.}
The first condition guarantees $n_s-1>0$ and the second condition guarantees $\ap_s<0$.
To see whether we can get a large running, let $f={\Delta'\over\Delta}\vp$, we have the
ratio
\eqn\rnwf{{(n_s-1)^2\over |\ap_s|}={(f+2)^2\over f^2+2f+f'\vp}.}
As a first trial, let $f=(\vp/M)^n$, then
\eqn\trone{{(n_s-1)^2\over |\ap_s|}={(x^n+2)^2\over x^{2n}+(n+2)x^n},}
where $x=\vp/M$.
To have a small ratio, $n$ must be large and $f=x^n\sim 2$. Indeed, take $f=x^n=2$,
the above ratio becomes $8/(n+4)$, and $n$ must be sufficiently large to get a value
close to $1/2$. For $f=2$, $n_s-1=4\Delta$, to have $n_s-1=0.2$, $\Delta =1/20$.
To be more concrete, for the choice $f=(\vp/M)^n$
\eqn\ptt{V\sim \Delta=\Delta_0\exp({1\over n}({\vp\over M})^n).}
For the choice $\Delta =20$, the dominance condition \dcc\ and the slow-roll
condition \nsc\ can barely be both satisfied. Using \numef\ we have
\eqn\efnum{N=\Delta_0^{-1}\int {d\vp\over\vp}exp(-{1\over n}({\vp\over M})^n),}
for $({\vp\over M})^n=2$ and smaller, the exponential function in the above
integral is a slowly-varying function, thus approximately we have
$N=\Delta_0^{-1}\ln{\vp\over\vp_*}$. Taking $\Delta_0^{-1}=20$, we find
$\vp=\vp_*e^3$ for $N=60$. Finally, we need to check whether the other slow-roll
condition $\eta\ll 1$ is satisfied. By definition of $f$, we have
$\eta=M_p^2/\vp^2 (f^2-f+f'\vp)$, which is just $M_p^2/\vp^2 (f^2+(n-1)f)$ for $f\sim
\vp^n$. As long as $M_p^2/\vp^2$ is sufficiently small, $\eta\ll 1$.

In general, from \rnwf\ we see that to have a small ratio $(n_s-1)^2/|\ap_s|$, $f'\vp$
must be larger than $f$. Now, $f=d\ln\Delta/d\ln\vp =d\ln V/d\ln\vp$, this condition
simply states that the second derivative of function $\ln V$ with respect to $\ln\vp$
must be greater than the first derivative. This eases the problem of large running
spectral index, since in the usual slow-roll inflation model, one typically requires
a large third derivative.

$\bullet$ Other models

Although a blue spectrum ($n_s>1$) is common in our nonminmal inflation scenario, it
is quite difficult to get a large running, as we shall show by considering some
examples.

\noindent $\diamondsuit$ Monomial potential

Let $\Delta=({\vp\over M})^n$,  to have a positive $n_s-1$, $n>-2$, since $f=\vp \Delta'
/\Delta=n$. To get a negative $\ap_s$, $n>0$. According to \rnwf, the ratio
$(n_s-1)^2/|\ap_s|=(n+2)^2/(n^2+2n)>1$.

The e-folds number is given by $N=(1/n)(\Delta^{-1}(\vp_*)-\Delta^{-1}(\vp))$. Apparently,
to have a large enough $N$, $\Delta (\vp_*)$ must be sufficiently small, thus, if $n\sim O(1)$,
$n_s-1$ is very small near the end of the inflation. However, $\Delta (\vp)$ does not need
to be very small, so it is still possible to have a quite blue spectrum at large scales.

\noindent $\diamondsuit$ Exponential potential

Let $\Delta =\Delta_0\exp({\vp\over M})$, we have
\eqn\exra{{(n_s-1)^2\over |\ap_s|}={(2+{\vp\over M})^2\over ({\vp\over M})^2+3{\vp\over M}}.}
To have a negative $\ap_s$ and $n_s>1$, $\vp/M>0$, thus the above ratio is again greater than 1.

\noindent $\diamondsuit$ Polynomial potential

Consider the special case $\Delta =\h ({\vp\over M})^2+{\ap\over n} ({\vp\over
M})^n$. Let $x=\ap(\vp/M)^{n-2}$, then
\eqn\ratt{{(n_s-1)^2\over |\ap_s|}={4+2({2\over n}+1)x+({2\over n}+1)^2x^2\over
2+({n\over2}+{4\over n}+1)x+({2\over n}+1)x^2}.}
Take a limiting case when $n\gg 1$, the above ratio assumes a minimal value when
$x\simeq 2$, and
\eqn\rattt{{(n_s-1)^2\over |\ap_s|}\simeq {12\over n+6},}
of course this value can be arbitrarily smaller than 1. This is a case with $f'\vp$ much larger
than $f$.

For $n=O(1)$, ratio \ratt\ can not be made small enough. We examine another limit in which
$n\ll 1$. Ratio \ratt\ assumes its minial value when
\eqn\minil{x={n\over 2}(\sqrt{3}-1),}
and \ratt\ is greater than 1. This is a special case of the following class of potentials.

\noindent $\diamondsuit$ Another class of polynomials

Take $\Delta =\Delta_0[1+({\vp\over M})^n]$. Let $x=({\vp\over M})^n$,
\eqn\rati{{(n_s-1)^2\over |\ap_s|}={(2+(n+2)x)^2\over n(n+1)(x+x^2)}.}
This ratio assumes its minimal value when $x=2/(n-2)$, and its value
is
\eqn\minv{{(n_s-1)^2\over |\ap_s|}={8\over n+1}.}
Once again, $n$ must be large to get a small ratio.

In conclusion, we have seen that the class of nonminimal inflation models we studied in
this note usually results in a blue power spectrum, and a large running of the spectral index
is possible, nevertheless the second derivative of the potential must be large enough.

\noindent Acknowledgments

The author is grateful to Jian Xin Lu, Tao Wang and Yi Wang for helpful
discussions. This work was supported by grants from CNSF.


\listrefs
\end